\documentclass[10pt,conference]{IEEEtran}

\ifCLASSOPTIONcompsoc
  \usepackage[nocompress]{cite}
\else
  \usepackage{cite}
\fi

\ifCLASSINFOpdf
\else
\fi
\usepackage{balance}  
\usepackage{graphics} 
\usepackage{txfonts}
\usepackage{times}    
\usepackage{color}
\usepackage{textcomp}
\usepackage{booktabs}
\usepackage{ccicons}

\usepackage[linesnumbered,noend, noline]{algorithm2e}
\usepackage{cite}
\usepackage{url}
\usepackage{fancybox}
\usepackage{multirow}
\usepackage{flushend}
\usepackage{booktabs}
\usepackage{tabularx}
\usepackage{comment}
\usepackage{array}
\usepackage[flushleft]{threeparttable}
\usepackage{mdframed}
\DeclareGraphicsExtensions{.pdf,.png}

\usepackage{relsize}
\usepackage{listings}
\usepackage{courier}

\usepackage{tikz}
\definecolor{1c1}{RGB}{188,162,6}
\definecolor{1c2}{RGB}{137,129,80}
\definecolor{1c3}{RGB}{239,167,31}
\definecolor{1c4}{RGB}{88,194,241}
\definecolor{1c5}{RGB}{6,180,188}
\tikzset{mynode/.style={draw=white,solid,circle,fill=green,inner sep=1pt, thick,
text=black}}
\tikzset{arrow line/.style={dashed, line width= 2.5pt, color=#1}}

\def\code{\tt}

\def\bf{\textbf}
\def\eq {Equation~}

\def\fig {Figure~}

\def\sec {Section~}

\def\alg {Algorithm~}

\def\it{\textit}
\def\tr{\textrm}
\def\tt{\mct}
\newcommand{\urls}[1]{{\scriptsize\url{#1}}}

\newcommand{\mct}[1]{{\footnotesize {\texttt {#1}}}}

\newcommand{\qu}[1]{{\it{``#1''}}}
\newcommand{\api}[1]{{\sf{\small{\texttt{#1}}}}}

\usepackage{paralist}

\usepackage{enumitem}

\newcommand{\nd}{\vspace{1mm}\noindent}
\usepackage[small,bf]{caption}

\usepackage{scalerel}

\usepackage{listings}
  \usepackage{courier}
\begin{document}

\title{Resolving API Mentions in Informal Documents}

\author{\IEEEauthorblockN{Gias Uddin and Martin P. Robillard}
\IEEEauthorblockA{School of Computer Science\\
McGill University\\
Montr\'{e}al, QC, Canada\\
Email: \{gias, martin\}@cs.mcgill.ca}
}

\IEEEtitleabstractindextext{%
\begin{abstract}
Developer forums contain opinions and information related to the usage 
of APIs. API names in forum posts are 
often not explicitly linked to their official resources. 
\it{Automatic} linking of an API mention to its official resources can be
challenging for various reasons, such as, name overloading. We present a
technique, ANACE, to automatically resolve API mentions in the textual contents 
of forum posts.
Given a database of APIs,  
we first detect all words in a 
forum post that are potential references to an API. 
We then use a combination of heuristics and 
machine learning to eliminate false positives 
and to link true positives to the actual APIs and their resources.  
\end{abstract}

}

\maketitle

\IEEEdisplaynontitleabstractindextext

%
\IEEEpeerreviewmaketitle

\section{Introduction}
Automatic traceability recovery between an API and its mentions 
in the forum posts can be useful to mine valuable information about the APIs. 
An API can be mentioned using its name (e.g., spring
framework), its code terms (e.g., \code{PropertiesFactoryBean}),
or hyperlinks to its resources (e.g., \urls{https://spring.io/}). The problem 
of resolving code terms in
API-related documents deals with tracing a code term to
its API (e.g., linking the type \code{PropertiesFactoryBean} to the API it
belongs to)~\cite{Dagenais-RecoDocPaper-ICSE2012a,
Rigby-CodeElementInformalDocument-ICSE2013,Subramanian-LiveAPIDocumentation-ICSE2014,Bacchelli-LinkEmailSourceCode-ICSE2010a}.
However, we are aware
of no technique to resolve mentions of general API \it{names} 
in the \it{textual} contents of the forum posts (see
\sec\ref{sec:background}).

We denote a phrase (e.g., spring) resembling an API name in a forum
post as a \it{named API mention}. We define the problem 
of \it{resolving such a named API mention} as
determining whether the mention actually refers to an
API and, if so,
which exact API it refers to.
We present a technique, ANACE (\bf{A}PI \bf{NAa}me Tra\bf{CE}rs), which, given a
database of APIs, detects API names in the forum posts and
links the names to their resource pages.   
First, we detect all API mentions, i.e., phrases in a 
forum post that are potential references to an API in our database.
We then use a combination of heuristics and 
machine learning to eliminate false positives 
and to link true positives to an actual API.

API names cannot be resolved 
with simple name-matching, when, for example, a mention can
match more than one API name. In fact, in our study of API
mentions we observed nine distinct sources of ambiguities in
API names that cannot be
resolved using trivial name matching (see \sec\ref{sec:ambiguity}).

Assigning a mention merely to the most popular API with the same name can also
be imprecise (e.g., most
 used API in Ohloh~\cite{website:ohloh-api} or downloaded in 
 Sourceforge~\cite{website:sourceforge}).
For example, such a strategy will always resolve a mention of `spring' or
`jackson' to their most popular API namesakes,
when the mentions may refer to other APIs or do not refer to any API at all
(e.g., jackson as a person or spring as a season).  
ANACE combines contextual information around an API mention 
with other features (e.g., contextual and structural
cues, API popularity) to determine correct resolutions.

\begin{figure}[t]
\centering
	\centering
   	\includegraphics[width=.48\textwidth,height=6cm]{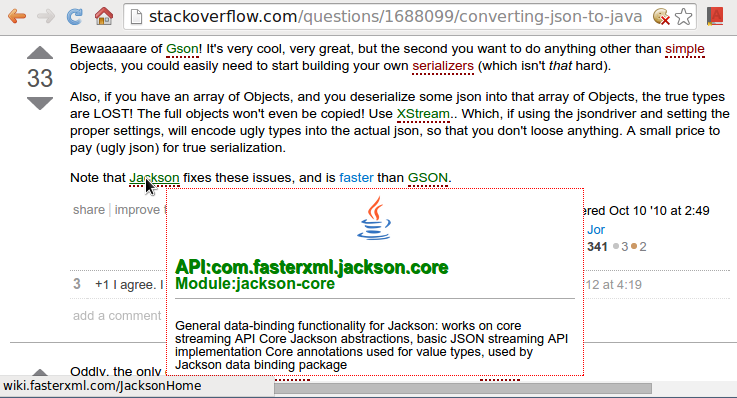}
   	\caption{The screenshot of ANACE }
 \label{fig:ANACE}
\end{figure}
In \fig\ref{fig:ANACE}, we show the screen shot of a client UI leveraging
ANACE for a StackOverflow thread. Each true mention is highlighted in green and false
   	ones in red. Each true
   	mention is assigned a link. For example, Jackson is resolved to
   	\api{com.fasterxml.jackson.core}. 
   	The bottom half of the tooltip shows a description of the API. 
 Clicking the mention
   	`jackson', leads to the API homepage (see the status bar).

\section{Ambiguity in API Mentions}\label{sec:ambiguity}
A \it{mention} in a post is a reference to an API. 
A mention can be one of the following
types:
\begin{inparaenum}[(1)]
  \item \bf{Name:} A name as a token (e.g., Jackson) or a series of
  tokens (e.g., Jackson JSON parser),
\item  \bf{Link:} A link to an API resource (e.g., homepage). 
\item \bf{Code:} Code (or code like) term/
snippet using packages and code elements from the API.
\end{inparaenum} We focus on the resolution of mentions referenced by name. We observed
nine sources of ambiguities in such mentions.

\begin{enumerate}
  \item \bf{\it{Homonymy:}} Multiple APIs can have
  the same name.
  \item \bf{\it{Meronymy:}} Instead of using an API name, developers may
  refer to its modules. Consider the following
  post~\cite{website:stackoverflow-q-2306190}: \qu{I'm building 
  my first real desktop applications\ldots I'm not sure if I should use SWT or
  Swing}.
  Here, `SWT' refers to the \api{SWT} module of the \api{Eclipse}
  framework. 
  \item \bf{\it{Synonymy:}} A single API can have more than one
  name. The \api{GSON} API can also be referred to as \api{google-gson}. 
  Both \api{org.glassfish.jersey} and \api{com.sun.jersey} refer to the same
  `jersey' framework.
	\item \bf{\it{Holonymy:}} A framework can interface with third-party
	APIs through dedicated modules. The \api{Apache camel} API offers
	integration with the \api{Jackson} API through its \api{camel-jackson} module.
	The module can still be referred to as: \qu{In
	apache camel, use Jackson for JSON parsing}. 
 
 \item \bf{\it{Hypernymy:}} An API from a given framework can be
 mentioned simply by the framework name. Due to the widespread
  usage of the JSON processor offered by the \api{com.fasterxml.jackson.core}
  project, the API is mostly referred to by simply `jackson'.

  \item \bf{\it{Spuriousness:}} This ambiguity is a special case of homonymy,
  where a mention that matches one or more API/module names may not
  refer to any of those. For example, 
  a mention
   of `jackson' is spurious if it refers to a non-API entity (e.g., a
   person).
   \item \bf{\it{Aliasing:}} A mention of an API is an alias 
   if it differs from the official name
   of the API. For example, 
   the \api{Google GSON} was mentioned as the `Google JSON' API. This
   is a special case of synonymy, where the synonym does not share any
   token with the name of the API it refers to (after removing stopwords 
   and organization names).
  \item \bf{\it{Demonymy:}} When the APIs implementing a specification share
  similar names with the specification, it is challenging to distinguish between
  the two. For example, the `dao' reference implementation in the
  \api{generic-dao} API.
  \api{Apache tomcat} is referred to both as a web server and as an open
  source API in the forum with the same name.
   
  \item \bf{\it{Platform-specificity:}} An API can have multiple versions to
  support different computing platforms. E.g., \api{org.json.me} is a
  mobile-optimized version of the \api{org.json} API, but it can still be
  referred to by \api{org.json} (see \cite{website:stackoverflow-q9009777}).
\end{enumerate}
We present techniques to resolve the first six ambiguities. 
The resolution of the other ambiguities is our future work.

\section{Related Work}\label{sec:background}
Related work can broadly be divided into three
categories:
\begin{inparaenum}[(1)]
\item
code term tracing,
\item
developer forum analysis,
and
\item
feature location.
\end{inparaenum}

\nd\bf{Code Term Traceability Recovery.}
Recodoc~\cite{Dagenais-RecoDocPaper-ICSE2012a} resolves code terms in the formal
documentation of a project to its exact corresponding 
element in the code of the project. 
Baker~\cite{Subramanian-LiveAPIDocumentation-ICSE2014}  
links code terms in the code snippets
of Stack Overflow posts to an API/framework whose name was used to tag the
corresponding thread of the post.
ACE~\cite{Rigby-CodeElementInformalDocument-ICSE2013} resolves Java code terms
in forum textual contents of posts using island
grammars~\cite{Moonen-IslandParser-WCRE2001}. Bacchelli et
al.~\cite{Bacchelli-LinkEmailSourceCode-ICSE2010a} developed Miler to determine 
whether a development email of a project contains the
mention of a given source code element. They compared information retrieval (IR)
techniques (LSI~\cite{Marcus-TraceabilityLSI-ICSE2003} 
and VSM~\cite{Manning-IRIntroBook-Cambridge2009}) against lightweight techniques
based on regular expressions. 
 Prior to Miler, LSI was also used by Marcus et al.~\cite{Marcus-TraceabilityLSI-ICSE2003}, 
and VSM was compared with a probabilistic IR model by Antoniol et
al.~\cite{Antoniol-TraceabilityCodeDocumentation-IEEETSE2002}. 
Tools and techniques have been proposed to leverage code traceability
techniques, e.g., linking software artifacts where a code
term is found~\cite{Inozemtseva-IntegrateSEArtifacts-ICSE2014}, 
associating development
emails with the source code in developers' IDE~\cite{Bacchelli-Remail-ICSM2011},
recommending posts in
Stack Overflow relevant to a given code context in the
IDE~\cite{Ponzanelli-PrompterRecommender-MSR2014}.

\nd\bf{Developer Forum Analysis} has been studied extensively,
e.g., to find dominant discussion
topics~\cite{Barua-StackoverflowTopics-ESE2012,Rosen-MobileDeveloperSO-ESE2015},
to analyze the quality of posts
and their roles in the Q\&A process~\cite{Calefato-SOSuccessfulAnswers-MSR2014,Bajaj-MiningQuestionsSO-MSR2014,
Lal-MigratedQuestionsSO-APSEC2014,Correa-DeletedQuestionSO-WWW2014,Vasilescu-SocialQAKnowledgeSharing-CSCW2014,Kavaler-APIsUsedinAndroidMarket-SOCINFO2013},
to analyze developer
profiles (e.g., personality traits of the most and low
reputed users) ~\cite{Bazelli-SOPersonalityTraits-ICSM2013,
Ginsca-UserProfiling-DUBMOD2013}, and to determine the influence of badges
in StackOverflow~\cite{Anderson-SOBadge-WWW2013}.
Tools have been developed using the knowledge
in the forums, such as,
autocomment assistance~\cite{Wong-AutoCommentSO-ASE2013},
collaborative problem
solving~\cite{Chang-RoutingQuestionsSO-SNAM2013,Tausczik-CollaborativeProblemSolving-CSCW2014},
and tag prediction~\cite{Stanley-PredictSOTags-ICCM2013}.

\nd\bf{Defect and Feature Traceability.}
Hayes et
al.~\cite{Hayes-TraceabilityAdvancingStudy-IEEETSE2006} used three IR
algorithms (LSI, VSM, and VSM with thesaurus) to establish links between
a high and low-level
requirement descriptions.
Lormans et al.~\cite{Lormans-TraceabilityLSIReconstructing-CSMR2006} used LSI to find
relationships between requirements, tests, and design documents.
Baysal et
al.~\cite{Baysal-CorrelateSocialInterationsReleaseHistory-MSR2007} correlated
emails in mailing lists and software releases by linking emails with the source
code. Types and variable names in the source code were matched against natural
language queries to assist in feature
location~\cite{PoshyvanykMarcus-LSIwithFCA-ICPC2007a,Eisenbarth-FeatureLocation-TSE2003,Marcus-TraceabilityLSI-ICSE2003}.
Given as input a bug report, Hipikat~\cite{CubranicGail-Hipikat-TSE2006} finds relevant source code and other artifacts (e.g., another bug report).
Subsequent techniques linked a bug fix
report to its related code
changes~\cite{Wu-RelinkBugChange-FSE2011,Nguyen-LinkBugReportFixes-FSE2012a},
or detected duplicate bug reports~\cite{Nguyen-DuplicateBugReport-ASE2012}.

\nd\bf{Discussion.}
To the best of our knowledge, no technique other than
ANACE exists to resolve \it{API names} in forum posts.
The code
term detection techniques rely on language syntax and 
naming conventions and thus cannot be adapted to detect API names,
because no
such structure exists for general API names. 
As explained in \sec\ref{sec:ambiguity}, the linking of an API
mention to an API is a multi-faceted problem due to diverse sources of
ambiguities. Such ambiguities do not come into play in the resolution of code
terms~\cite{Dagenais-RecoDocPaper-ICSE2012a, Rigby-CodeElementInformalDocument-ICSE2013,Subramanian-LiveAPIDocumentation-ICSE2014,Bacchelli-LinkEmailSourceCode-ICSE2010a}.
Similar to the code traceability techniques, ANACE also needs a 
pre-defined dictionary of entity names. 
Unlike Recodoc~\cite{Dagenais-RecoDocPaper-ICSE2012a} that operates on formal
documents, ANACE resolves API names in \it{informal documents}. 
Both Baker~\cite{Subramanian-LiveAPIDocumentation-ICSE2014} and ACE~\cite{Rigby-CodeElementInformalDocument-ICSE2013}
assume a semi-open scope by relying on tags to 
filter out posts that may not represent an API name of interest. 
We take an \it{open scope} 
by assuming that a thread can contain discussion about any API.

\begin{algorithm}[t]
 \SetKwInOut{Input}{input}\SetKwInOut{Output}{output}
 \SetKwFunction{add}{append}
 \SetKwFunction{resolve}{getClassifyConf}
 \SetKwFunction{filter}{filter}
 \SetKwFunction{name}{name}
 \SetKwFunction{getHomepage}{getHomepage}
  \SetKwFunction{len}{length}
 \SetKw{KwNext}{next}
 \SetKw{break}{break}
 \SetKw{null}{null}
 \Input{
 \begin{inparaenum}[(1)]
 \item Mention Candidate List, $MCL$,
 \item Trained resolution classifier \tt{RC} 
 \end{inparaenum}
  
 }
 \Output{Resolution decision, $D = (d_{api}, d_{module}, d_{url})$
 }
 $H = \emptyset$,  	$d_{module}$ = \null, $d_{url}$ = \null\;
 	\ForEach{\tr{candidate $c_i$ $\in$  $MCL$}} {
 		$confidence$ = \resolve{$mention, c_i$}\;
 		\lIf{$confidence > \tau$}{$H$ = $H\cup \{c_i\}$}
 	}
 	\lIf{$|H|$ = $0$}{
 		$D$ = $\emptyset$, \KwRet $D$
 	}
 	\lElseIf{$|H|$ = 1}{
 		$H = \{c_i\}$, $d_{api} = c_i$
 	}
 	\lElse{$d_{api} = \filter{$H$}$
 	}

 	\ForEach{module $s_i$ $\in$ $d_{api}$}
 	{
 		\lIf{\tr{Mention} = \name{$s_i$}}{$d_{module}$ = $s_i$, \break}
 	}
 	\If{$d_{module}$ $\ne$ \null}{
 		$d_{url}$ = \getHomepage{$d_{module}$}\;
 	} 
 	\lElse
 	{
 		$d_{url}$ = \getHomepage{$d_{api}$}
 	}
 $D = (d_{api}, d_{module}, d_{url})$,
 \KwRet $D$\;
 \SetKwProg{proc}{procedure}{}{}
 \proc{\resolve{$m,c$}}{
 	\KwRet classify ($m,c$) using \tt{RC}\;
 
 }
 \proc{\filter{$H$}}{
 	\KwRet a candidate $c$ in $H$ using filters\;
 }
 \proc{\getHomepage{$c$}}{
 $d_{url}$ = most frequent url in $c$, 
 \KwRet $d_{url}$\;
 }
 \caption{The resolution of a mention to an API}
 \label{alg:resolution}
\end{algorithm}

\section{Resolution Framework}\label{sec:framework}
Our API database consists of the Java official APIs and the open
source Java APIs. 
Each entry in the database
contains a reference to a Java API. 
For each API, we collect seven fields from online portals:
\begin{inparaenum}[(1)]
\item API name 
\item module names
\item resource links, e.g., download page, documentation page, etc.  
\item overview description
\item license and organization information 
\item dependency on another API
\item usage count: Every project page in Ohloh 
shows how many Ohloh users listed the API in their
personal development stack
\item download count: If the API is also hosted in Sourceforge, we 
collect how many times the API was
downloaded.
\end{inparaenum}

ANACE operates in four steps: 
\begin{inparaenum}[(1)]
\item We \bf{crawl} the online software portals to create
the API database and forum posts. 
\item We \bf{preprocess} the contents.
\item We \bf{detect} phrases in the forum contents that
match at least one API/module name in the database. Each such phrase is called
an API \it{mention}, which can match more than one API name. Each such match
is called a \it{candidate}.
\item We \bf{resolve} an API mention to only one
candidate, or label it as false.
\end{inparaenum} 

\begin{figure}[tbp]
  \centering

  \includegraphics[scale=0.50]{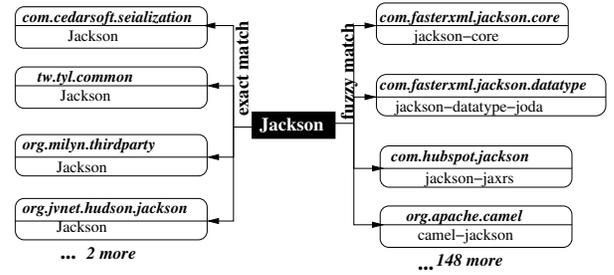}
  \caption{Partial Mention-Candidate List (MCL) for `Jackson'.}
  \label{fig:mc-graph}
  \vspace{-.8cm}
\end{figure}

\nd\bf{Mention Detection.} We consider a token (or a series of tokens) in a
forum post as a mention if it matches at least one API or module name in 
our API database. We use both exact and fuzzy name
matching (see \sec\ref{subsec:name-similarity}). For each detected mention, 
ANACE produces
a Mention Candidate List (MCL) as follows:
  Match the token(s) against all the APIs in the database.
  Return as a potential candidate an API whose name matches
  (exact/fuzzy) the token(s), or
  return as a potential candidate an API if at least one of its module
  names matches the token(s). Hence, an MCL contains a mention, linked to one or
  more candidates from the API database. 
  In \fig\ref{fig:mc-graph},
we show a partial mention candidate list for the mention 
`Jackson' shown in \fig\ref{fig:ANACE}. Each rounded rectangle
denotes an API candidate with its name at the top and module names at
the bottom (if module names matched). 

\nd\bf{Mention Resolution (\alg\ref{alg:resolution}).} The
process has two steps:
\begin{inparaenum}[\bfseries(1)]
  \item Given a mention-candidate list, we resolve the
  mention to one of its candidates (e.g., jackson 
  to \api{com.fasterxml.jackson.core} in \fig\ref{fig:ANACE}).
  \item Given a resolved API, we
  assign a resource link to the mention (e.g.,
  \url{http://wiki.fasterxml.com/JacksonHome}).
\end{inparaenum} 
\alg\ref{alg:resolution} takes as input a mention candidate list and resolution
classifier (\tt{RC}). The classifier produces a confidence value
($[0, 1]$) between the mention and each candidate: A
value of 1 denotes that it has \it{complete} confidence
that the mention can be resolved to the candidate. 
The classifier is supervised, which we trained on our development
dataset (see \sec\ref{sec:linking}). A confidence value for a candidate above
0.5 (i.e., $\tau$ in line 4) is considered as a
\it{hit}, i.e., it could be a probable resolution.
For a mention-candidate list, the classifier may find more than one hit.
If the list of hits is empty, we label the mention as false.
For only one hit, we resolve the mention to the hit.
For more than one hit, 
 we apply two types of filters to decide which of the hits is the correct
 resolution (see \sec\ref{sec:filters}).

\section{The Resolution Classifier}\label{sec:linking}
We used a Na\"{i}ve Bayes classifier (\tt{RC} in
\alg\ref{alg:resolution}) to calculate a confidence value for each candidate in
a mention candidate list.
We compute three types of similarity weights (range $[0,1]$) between the mention 
and each candidate: name
(see \sec\ref{subsec:name-similarity}), context
(\sec\ref{subsec:similarity-measures}), and structural (\sec\ref{subsec:sim-structural}).
To produce the confidence value for a candidate, the classifier uses its
similarity weights and two popularity counts: its usage and download counts.

\subsection{Name Similarity}\label{subsec:name-similarity}
A name similarity weight greater than 0 between a
mention and an API or its module name was used to include the API in the mention
candidate list. An \bf{exact match} between a
mention (M) and a candidate (C) API/module name is
considered only if both contained the same series of tokens in the same order.
We assigned the similarity a weight of 1.
For \bf{fuzzy matching}, we employ two techniques:
\begin{inparaenum}[(1)]
\item \it{Prefix Matching} is defined as $M$ and $C$ both sharing the same
prefix. The weight is 1.
\item \it{Token Sorting} is defined as $M$ and $C$ both having one or
more shared tokens, e.g., $M$ = `jackson' and $C$ = \api{com.fasterxml.jackson.core}.
The similarity weight is the Jaccard
index~\cite{Manning-IRIntroBook-Cambridge2009}:
\begin{equation}\label{eq:jaccard-match}
w = \frac{|\tr{Tokens}(M)\cap \tr{Tokens}(C)|}{|\tr{Tokens}(M)\cup
\tr{Tokens}(C)|}
\end{equation}
\end{inparaenum}

\subsection{Context Similarity}\label{subsec:similarity-measures}
We analyze the text around a mention to construct a \it{feature context}. 
We compute context similarity by comparing 
the feature context against the description of each 
candidate. We compute two types of similarity: noun and
verb-based.

\nd\bf{Constructing The Feature Contexts for Mentions.} 
The feature context is a bag of tokens. 
We observed that when we find a mention 
in more than one post of the same thread, all
of those occurrences usually referred to one single API. We include
the following tokens in the feature context of a mention:
\begin{inparaenum}[(1)]
  \item \bf{same post}: tokens around it within a \it{window}.
  A window size of 3 takes tokens from 3 sentences right and 3 left (when available).
  \item \bf{other posts}: tokens within the window of same mention in other
  posts.
  \item \bf{title}: tokens in the title.
\end{inparaenum}

\nd\bf{Constructing Descriptions for Candidates.}
The description for each candidate is
a bag of tokens except stopwords from \it{selected} sentences
from its description found 
\begin{inparaenum}[(1)]
  \item in the portal, and
  \item in its homepage.
\end{inparaenum} 
Consider the description of the API \api{com.fasterxml.jackson.core} on its
homepage (denoted by $d_{H}$ afterwards):
\qu{Jackson is a high-performance json processor. It
provides a json parser\ldots This will be the portal page for Jackson}($d_{H}$).
The description of the API in our database as extracted from the portal (denoted
by $d_{P}$):
\qu{\ldots It provides \ldots Add-on module \ldots 
to support Joda
(\urls{http://joda-time.sourceforge.net/}) data types\ldots} ($d_{P}$). From the
descriptions, we only include a sentence if it:
\begin{inparaenum}[(1)]
\item starts with the name of the API or its module (e.g., 
\qu{Jackson is a high-performance JSON processor ...} (in $d_{H}$))
\item contains a subject pronoun referring to the API,
and the sentence immediately follows a sentence of type 1 (e.g., 
\qu{It provides a JSON parser \ldots}. (in $d_{H}$))
\item contains a reference to another API (e.g., 
\qu{Add-on module for Jackson to support Joda
(\urls{http://joda-time.sourceforge.net/}) data types.} (in $d_{P}$). Here, Joda
is a reference to \api{joda-time} API.
\end{inparaenum}

We consider a 
link or name as a reference to another API if:
 \begin{inparaenum}[(1)]
 \item the link refers to the resources of another API, or
 \item the name is in the list of dependencies of the API. 
 \end{inparaenum} We consider selected sentences based on our
observation that not all the sentences are essential to
learn about the features of a candidate. For example, the $d_{H}$ above also
contains:
\qu{This will be the portal page for Jackson project}.

For {Noun-based Similarity.}, we compute how the tokens tagged as nouns 
in the context
of a mention (M) are similar to the tokens tagged as nouns in the description of 
each of its
candidates using \eq\ref{eq:jaccard-match}. \bf{Verb-based Similarity} uses
the same approach, but analyzes the verbs instead of the nouns. 
 
\subsection{Structural Similarity}\label{subsec:sim-structural}
We heuristically link the code terms around a mention to
its candidates. The more code terms are \it{associated} with a candidate, the
more structurally it is similar to the mention.

\nd\bf{Constructing Code Context for Mentions.} We identify types (class,
interface) in each post using Java naming conventions,
similar to previous approaches~\cite{Dagenais-RecoDocPaper-ICSE2012a,
Rigby-CodeElementInformalDocument-ICSE2013} (e.g., camel case, etc.). 
We collect types that are most likely not declared by the user.
Consider the following example~\cite{website:StackOverflow-q-4486787}:

\begin{lstlisting}
import com.fasterxml.jackson.databind.*
private void tryConvert(String jsonStr) {
    ObjectMapper mapper = new ObjectMapper();  
    Wrapper wrapper = mapper.readValue(...);}
\end{lstlisting}
We add \code{ObjectMapper} into our code context, but not
the type \code{Wrapper}. This is because the same post later declares the
type \code{Wrapper} as: \code{public Class Wrapper}. We parse code snippets
using the ANTLR parsing
framework~\cite{Parr-ANTLR-Book2007}. 
We discard two types of snippets that the
ANTLR Java parser cannot parse:
\begin{inparaenum}[(1)]
\item Non-java snippets (e.g., .NET), and 
\item Malformed Java snippets (e.g., a mix of Java and XML, etc.).
\end{inparaenum} 

\begin{figure}[t]
  \centering

  \includegraphics[scale=0.65]{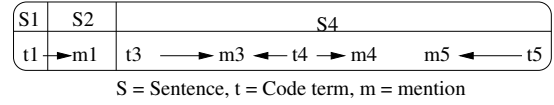}
  \caption{Assignment of code terms to mentions in a post. We process the
  sentences in sequence and assign a code term to its nearest mention.
  A mention is the nearest if (1) it is immediately after or before the code
  term in the same sentence,
  or (1) it is in a different sentence of the code term, but no other
  mention is found before.}
  \label{fig:struct-sim-window}
\end{figure}

In a post with only one
mention, we assign all types in the post to the code context of the
mention. In the presence of multiple mentions in the same post, we define a
\it{window} to assign types (explained in \fig\ref{fig:struct-sim-window})

\nd\bf{Linking Types to the Candidates
(\alg\ref{alg:struct-linking}).} The input is a type name found
in the code context of a mention, its candidate APIs, and the code snippets
from the same post. The output is a one or more candidate APIs to which the
type may belong to. If the type name is fully-qualified (FQN) (e.g.,
\code{com.fasterxml.jackson.databind.ObjectMapper}), 
we associate it to the candidate whose type name matches it exactly (line 4).
For an unqualified type name in the code context (e.g., \code{ObjectMapper}), 
we analyze the \code{import} statements (when
available) in 
the input code snippets (lines
7-9).
For example, the above code snippet imports the package
\code{com.fasterxml.jackson.databind.*} from the API \api{..jackson.core}.
There is an FQN in \api{..jackson.core} by the name
\code{..jackson.databind.ObjectMapper}. We thus associate
\code{ObjectMapper} to only \api{..jackson.core}. In the absence of import
statements, we associate the type to all of the APIs whose
type names (unqualified) matched the type (lines 6, 10).
We compute the
structural similarity between a mention $M$ and a candidate $C$ as:
\begin{equation}
\tr{simscore(structure)} = \frac{|\tr{Types}(M)\bigcap
\tr{Types}(C)|}{|\tr{Types}(M)|}
\end{equation} $\tr{Types}(M)$ is the list of types for $M$ in its context.
When code terms are not found for a given mention, 
we assign $\tr{Types}(M) = \emptyset$, i.e.,$|\tr{Types}(M)\bigcap
\tr{Types}(C)| = 0$ . 
\begin{algorithm}[t]
 \SetKwInOut{Input}{input}\SetKwInOut{Output}{output}
 \SetKwFunction{add}{append}
 \SetKwFunction{FQN}{FQN}
 \SetKwFunction{name}{UnqualifiedName}
 \SetKwFunction{isFromImported}{isInImported}
  \SetKwFunction{getProcessed}{getProcessed}
 \SetKw{KwTrue}{true}
 \SetKw{break}{break}
 \SetKw{null}{null}
 \Input{
 \begin{inparaenum}[(1)]
 \item Mention Candidate List, $MCL$ = $\{c_1, \ldots, c_n\}$,
 \item A type name \tt{T} from a code context, \item All code snippets $S$ in
 the post.
 \end{inparaenum}
  
 }
 \Output{Linking decision $D_T = \{c_i, \ldots\}$}
 $D_T = \emptyset$, $H [c_1] = \emptyset$ , $\ldots$, $H[c_n] = \emptyset$, $A = \emptyset$\;
 
 	\ForEach{\tr{candidate $c_i$ $\in$  $MCL$}} {
 		\ForEach{type $t_i$ $\in$ $c_i$}
 		{
 			\lIf{\FQN{$t_i$} = $T$}{$D_T = D_T \cup \{c_i\}$}
 			\ElseIf{\name {$t_i$} = $T$}
 			{
 				$H[c_i] = H[c_i] \cup \{t_i\}$, $A = A \cup \{c_i\}$\; 
 			}
 			
 		}
 	}
 	\ForEach{candidate $c_i$ $\in$ $H$}
 	{
 		\ForEach{type $t_i$ $\in$ $H[c_i]$ }
 		{
 			\If{\isFromImported{$T$, $t_i$}}{$D_T = D_T$ $\cup$ \{$c_i\}$\;}
 		}
 	}
 	\lIf{$|D_T| = 0$}{$D_T = A$}
  \KwRet $D_{T}$\;
  \SetKwProg{proc}{procedure}{}{}
 \proc{\isFromImported{$T$, $t_i$}}{
 	\ForEach {snippet $s_i \in S$ }
 	{
 		\ForEach{Import statement $i$ $\in$ $s_i$}
 		{
 			$T = \getProcessed{$i$} + '.' + T$\;
 			\lIf{$t_i = T$}{\KwRet \KwTrue}
 		}
 	}
 }
  \proc{\getProcessed{$i$}}{
  \lForEach{$t$ $\in$ $\{import , ; , *\}$}
  {
  	remove $t$ from $i$
  }
  \KwRet $i$\;
  }
 
 \caption{The linking of a type name to a candidate}
 \label{alg:struct-linking}
\end{algorithm}

\section{Candidate Filtering Heuristics}\label{sec:filters}
We considered candidates
with a confidence value $>$ 0.5 from the resolution classifier as potential
hits.
We observed that it can be erroneous to trivially select the candidate with the
highest confidence value because more than one candidate or their extension can
offer similar features, and the description of the most likely candidate are insufficient
  or incomplete.
We apply the following three filters in sequence as listed below to pick the
best hit. We do not apply a second intrinsic filter if the mention is already
resolved using another filter.

\begin{inparaenum}[\bfseries 1.]
\item\bf{Betweenness}: We apply this filter, if the feature context of the
mention contains the keywords `extension', `wrapper', `plugin', 
or variants thereof (e.g., `plug-in'). We determine whether 
a candidate $c_1$ in the hit-list is a direct extension 
 of another hit $c_2$ (i.e., direct incoming edge from $c_1$).
If so, we put the extension (i.e., $c_1$) into a bucket. 
Consider the sentence: \qu{Use the gson extension easy gson \ldots}.
Given a hit-list for the mention `easy gson' with two candidates (\api{gson}
and \api{easy-gson}), we put \api{easy-gson} in the bucket, because
it depends on \api{gson}. For only one candidate in the bucket, we assign
the mention to the candidate. For more than one
extension, we select the one with the highest name similarity. 

\item\bf{Centrality:} We compute the \it{influence} of
each candidate in a hit-list on the rest of the candidates in the same mention
candidate list.
For the mention `jackson' in the sentence \qu{I use Jackson to parse JSON
messages} (see
\fig\ref{fig:mc-graph}) and given a hit-list with two APIs
(\api{com.fasterxml.jackson.core} and \api{..jackson.datatype}), we compare which one of the two candidates is
used the most by the other candidates in the mention candidate list.
We compute the influence of a hit on other candidates by taking the proportion
of the number of other APIs in the mention candidate list that are dependent on
 a hit versus the number of other APIs that the hit is dependent on.  
We
assign the mention to the hit with the highest influence score. 
If we have
ties for the highest score, we assign the mention 
to the hit on which most other APIs are dependent on. Otherwise, we move to the
next filter.

\item\bf{Closeness}. We compute how close a hit of type `core' is with
other candidates in the mention candidate list and assign the mention to the hit
with the lowest `closeness' value.
\begin{equation}\label{eq:closeness-centrality}
\tr{Closeness (c)} = \frac{1}{\tr{\# Other APIs in MCL dependent on (c) +
1}}
\end{equation}
 The constant $1$ is used as a smoothing value, loosely
 adapted from the definitions of Laplace
 smoothing~\cite{Manning-IRIntroBook-Cambridge2009}. 
\end{inparaenum}

\subsection{Extrinsic Filters}
We determine whether and how
a mention relates
to the surrounding other mentions in the same post. We apply 
the following three filters in sequence: composition, aggregation, and
projection.
We stop processing a mention if we can make a decision using a filter. For a
given forum post, the input to each filter is a list of all hit-lists and the
mentions found in the post as produced by the resolution classifier, even when
the mentions may already be resolved by the intrinsic filters. 
If we can
make a decision using the extrinsic filters, but the mention is already
resolved through intrinsic filters, we overwrite the previous
decision (explained below).
Therefore, for these filters to be
applicable to a mention, we require at
least one true mention immediately preceding and one 
following the mention in the same post already resolved.  

\begin{inparaenum}[\bfseries 1.]
\item \bf{Composition:} We determine whether the candidate API can be a
  module of an API mentioned immediately before the candidate. 
  For the mention `jackson' in the sentence \qu{In apache-camel, Jackson can
  deserialize JSON}, suppose the hit-list includes two candidates:
\api{com.fasterxml.jackson.core} and \api{org.apache.camel}. We assign
`jackson' to \api{org.apache.camel} because 
one of its module named as \api{camel-jackson} 
which offers JSON processing features and the previous mention
\api{apache-camel} was resolved to the API \api{org.apache.camel}. 
For the above hit-list, the influence intrinsic filter
will erroneously assign the mention `jackson' to the \api{..jackson.core},
because \api{..apache.camel} depends on it. By applying this filter, we overwrite
the resolution to \api{..apache.camel} API.  

\item \bf{Aggregation:} We determine whether the immediately preceding or
following other mentions are dependent on the candidate. For the mention
`jackson' in ``Since spring packages Jackson, we used JSON-based messages'',
and a hit-list with \api{com.fasterxml.jackson.core} and
\api{..jackson.datatype}, we assign jackson to \api{..jackson.core}, 
because the nearest mention
to Jackson in the post is Spring, which is resolved to 
\api{org.springframework} and depends on \api{..jackson.core}. 

\item \bf{Projection:} We determine whether the candidate API is dependent on
any of the surrounding mentions in the same post. Consider the sentence:
``I can serialize Joda-time with the Jackson JSON processor''. Given 
\api{com.fasterxml.jackson.core} and \api{..jackson.datatype} as hits, 
we assign jackson to \api{..datatype}, because it depends on the
\api{joda-time} API.
\end{inparaenum}

\section{Summary}\label{sec:summary}
The resolution of API names in the developer forums
can be challenging when a mention can exhibit ambiguities, 
e.g., more than one API exist with the same name. 
We presented ANACE, a technique 
that can resolve API mentions in forum posts. Our ongoing work
focuses on the following directions:
\begin{itemize}
\item \bf{Evaluation}: The effectiveness of ANACE compared to the baselines
(eg., search engines, etc.)
\item \bf{Empirical Study}: Analysis of the prevalence of the
ambiguities in the forum post.
\item \bf{Extension}: The extension of ANACE to handle API name resolution from
different other programming languages.
\end{itemize}
\begin{small}
\bibliographystyle{IEEEtran}
\bibliography{consolidated}
\end{small}
\end{document}